\documentclass[aps]{revtex4-2}
\usepackage{hyperref}
\usepackage {chngcntr}
\usepackage{graphicx}
\usepackage{amsmath,mathtools}
\usepackage{latexsym}
\usepackage{amsfonts}
\usepackage{amssymb}
\usepackage{array}
\usepackage{color}
\usepackage{subfigure}
\usepackage{verbatim}
\usepackage{ulem}
\newcommand{\ket}[1]{\left\vert#1\right\rangle}
\newcommand{\bra}[1]{\left\langle#1\right\vert}
\newcommand{\av}[1]{\langle#1\rangle}

\newcommand{\ii}{\mathrm{i}}
\newcommand{\ee}{\mathrm{e}}

\newcommand{\A}{\alpha}

\definecolor{airforceblue}{rgb}{0.0, 0.0, 0.78}

\newcommand{\new}[1]{{\color{black}#1}}
\newcommand{\old}[1]{{\color{black}#1}}

\newcommand{\cmmnt}[1]{}
\begin{document}
\title{Efficient quantum simulation of  nonlinear  interactions using SNAP and Rabi gates}

\author{
Kimin Park, Petr Marek and Radim Filip}
\affiliation{
Department of Optics, Palacky University, 77146 Olomouc, Czech Republic}
\date{\today}

\begin{abstract}

\new{Quantum simulations provide means to probe challenging problems within controllable quantum systems.}
However, \new{implementing or simulating} deep-strong nonlinear couplings between bosonic oscillators on physical platforms remains \new{a challenge}.
\new{We present a deterministic simulation technique that efficiently and accurately models nonlinear bosonic dynamics. 
This technique alternates between tunable Rabi and SNAP gates, both of which are available on experimental platforms such as trapped ions and superconducting circuits.}
Our \new{proposed simulation method facilitates} high-fidelity \new{modeling} of  phenomena that emerge from higher-order bosonic  interactions, \new{with an exponential reduction in resource usage compared to other techniques}.
\new{We demonstrate the potential of our technique by accurately reproducing key phenomena and other distinctive characteristics of ideal nonlinear optomechanical systems.}
\new{Our technique serves as a valuable tool for simulating complex quantum interactions, simultaneously paving the way for new capabilities in quantum computing through the use of hybrid qubit-oscillator systems.}

\end{abstract}

\maketitle




\section{Introduction}

Advances in the strong coupling regime of qubit-oscillator interactions in trapped ions and superconducting circuits \cite{Forn-DiazRMP2019Rabi,NoriRabiRMP2019,VijayNature2012Rabi} has opened \new{new revenues for} building qubit gates \cite{Devoret2013ScienceSuperconducting,TanNature2015TrappedIon,MilnePRApp2020PMgate} and, mainly, controllable gates of the oscillators \cite{FluhmannHomeNature2019,PuriSciAdv2020Cat,GrimmNature2020KerrCat} and first nonlinear gates \cite{YoshiharaNatPhys2017Ultrastrong,HeeresNatComm2017,GaoPRX2018}. 
Trapped  ions~\cite{BruzewiczAPR2019TrappedIon,wolf2019NatCommFocktrappedion,KatzPhysRevLett2023SimBosTrappedIon} and superconducting circuit QED with transmons~\cite{BlaisRMP2021Supercond,Gu2017,Ofek2016Nature} are highly promising platforms for qubit-oscillator quantum information processing.
The fundamental qubit-oscillator interaction is the resonant Rabi coupling\old{~\cite{Forn-DiazRMP2019Rabi,NoriRabiRMP2019}}, while dispersive interaction appears  far from resonance \cite{Zueco2009,BlaisRMP2021Supercond}.   
Strong Rabi couplings have been  experimentally demonstrated in cavity QED systems, including trapped ions~\cite{Lv2018RabiTrappedIon} and superconducting circuits~\cite{Markovic2018RabiSC}.
These enable diverse non-Gaussian operations such as nonlinear phase gates~\cite{ParkPRA2016, ParkNJP2018}, which are essential for CV quantum information processing~\cite{WeedbrookRMP2012CV}.
\new{Both trapped ion and superconducting circuits can dynamically switch between interaction types.
In particular, superconducting circuits show promise for implementing tunable dispersive and resonant qubit-oscillator couplings~\cite{Zueco2009,GaoPRX2018,Eickbusch2021SupConDispersive,SchoelkopfNature2020,StehlikPRL2021Tunable,PanPRX2022}, whereas achieving dispersive coupling remains challenging in trapped ion systems~\cite{FluhmanPhysRevLett2020Trappedion}}.
In this regime beyond the rotating-wave approximation, \new{theory suggests that we can achieve} universal control of the oscillator using sequences of Rabi gates \cite{ParkNJP2018} \old{and dispersive gates}~\cite{KrastanovPRA2015dispersive}. 
Such methodology offers quantum simulations of nonlinear effects with general bosonic systems \cite{LiPhysRevApplied2018StateTransfer,LangfordNatComm2017Rabi}. 
\old{Recently, Kudra et al. \cite{kudraPRXQ2022robust} experimentally generated high-fidelity Wigner-negative states, including Schrödinger-cat states and Gottesman-Kitaev-Preskill states, \old{using novel SNAP gates} in 3D microwave cavities.}
It is \new{intriguing to consider} whether such a sequence of pulsed gates   can construct \new{a variety of} nonlinear interactions between oscillators and efficiently simulate more complex hybrid nonlinear phenomena in quantum physics. 


A \new{prototypical} example is \old{the strong nonlinear radiation pressure interaction between a movable mirror in a Fabry-Perot cavity and the intracavity field \cite{AspelmeyerRMP2014,Millen2019} in nonlinear optomechanics. }
This interaction is fundamentally nonlinear, involving \old{third power of} ladder operators in the Hamiltonian~\cite{AspelmeyerRMP2014}.
At the quantum level, \old{such} a resulting \old{pressure-like} interaction enables\old{, in principle,} \old{challenging} single-photon-scale displacements of the \old{mirror}. 
Exploring this highly nonlinear coupling, which can generate nonclassical superposition states from phase-insensitive noise\old{~\cite{oconnell2010Nature,wollman2015Science,ockeloen2018Nature,clark2016NatPhy}}, will substantially advance current   quantum technologies \cite{Shkarin2019,Wiederhecker2019,PiergentiliNJP2018,Tavernarakis2018,LombardiPRX2018,Renninger2018,DelicQST2020Dispersive,JohnssonPRR2021} and therefore, \old{it is} attractive for hybrid quantum simulations~\cite{GeorgescuRMP2014}. 
Nonlinear \old{pressure-like} phenomena are predicted to enable breakthrough applications in quantum information processing, metrology, communication, fundamental tests of quantum mechanics~\cite{StannigelPRL2010,MarinkovicPRL2018Bell,PikovskiNatPhys2012}.

However, the inherent weakness of \old{nonlinear radiation pressure} interactions \new{presents a challenge in directly observing} quantum nonlinear \old{phenomena}  without \new{the introduction of} \old{additional} noise.
Although strong optical pumping can enhance  effective coupling strengths,  it inevitably leads to linearization of the coupling~\cite{AspelmeyerRMP2014}, \old{removing} the underlying nonlinear dynamics. 
Experimental strategies like amplification have also been employed to access the single-photon scale nonlinear optomechanical phenomena~\cite{xgd12, rbap14, hmtks14, pcmthhs15, lwjjzn15, ldc16, ws17}.
The membrane-in-the-middle configuration provides enhanced nonlinear optomechanical interactions~\cite{DumontOptExp2019Membrane}, making it an attractive resource for quantum simulation of the hybrid system.
Nevertheless, proof-of-principle demonstrations and analysis of these higher-order nonlinearities remain challenging, \old{and finding new routes to harness these interactions has been an attractive prospect for quantum technologies}. 

An effective \new{strategy} for \new{exploring} deep-strong \old{and nonlinear} quantum  phenomena \new{involves the use of} quantum simulation \new{based on} engineered interactions~\cite{Hartmann2016nonlinearsimulator}. 
There are several methods available for engineering continuous-variable gates~\cite{KalajdzievskiQuantum2021CVDecomp}.
One such method involves simulating a conditional form of optomechanical interaction~\cite{ParkPRA2015Optomechanics} using photonic ancillas and homodyne detectors~\cite{ParkPRA2014Xgate}.
Despite its potential, the achievable coupling strength is weak, and the implementation is probabilistic.
\new{Previous methods \cite{JacobsPRL2007, ParkSciRep2017Kerr,ParkNJP2018} to engineer operations using an auxiliary qubit mainly focused on higher-order single-mode Hamiltonians. 
However, extending these to two-mode simulations presented significant complexities, particularly the necessity of an exotic three-body interaction for   \cite{JacobsPRL2007}.}
Recently, schemes utilizing the optical Fredkin gate \old{and cross-Kerr interaction} induced by a strong drive have been proposed~\cite{LiaoPRA2020OptoMechKerrSim,Yin2021, ZhouOptExp2021Simulation}.
However, these methods \new{involve linearizing} a naturally weak, higher-order optical interaction  into a lower-order interaction, \new{leading to decreased}    accuracy and resource efficiency. 
Extending these methods to simulate various \old{nonlinear} couplings, especially between \old{hybrid} modes \old{at the different platforms}, is challenging.
In this work, we introduce a highly efficient constructive approach 
\new{to simulate} the dynamics \new{of} deep-strong quantum multimode bosonic coupling.
We show how alternating a selective dispersive interaction conditioned on the individual Fock state in a cavity mode (\old{SNAP gate})~\old{\cite{SchusterNature2007, HeeresPRL2015}} with Rabi gates can effectively reproduce arbitrary-order radiation-pressure interaction.
Our approach does not rely on existing nonlinearities between oscillators and offers  an exponential enhancement in the resource usage over other methods, reproducing distinctive effects of radiation pressure and membrane-in-the-middle optomechanical systems.
Superconducting circuits \new{are} a promising platform for realizing this approach.
Recent advancements have demonstrated the \new{feasibility of coupling} a transmon qubit to multiple cavity modes and \new{achieving} switchable dispersive and resonant interactions \cite{Eickbusch2021SupConDispersive,SchoelkopfNature2020,StehlikPRL2021Tunable,PanPRX2022}.
Our work \new{not only facilitates the exploration of experimentally challenging nonlinear coupling regimes but also significantly enhances the capabilities of quantum simulation}, opening up new capabilities for hybrid quantum information processing using qubits and oscillators and tests of new quantum nonlinear phenomena.



\new{\section{Method}}
\label{sec:int}

Our \new{objective} is \old{to implement  general deep-strong cavity optomechanical} Hamiltonians, \new{ denoted as}
$\hat{H}_{k,l} =  \Omega_{k,l} \hat{n}_L^k \hat{X}_M^l$.
\new{Here} $\Omega_{k,l}$ \new{represents} the coupling strength, and $k,l\in \mathbb{Z}$. 
Here, the number operator  $\hat{n}_L=\hat{a}^\dagger\hat{a}$ and a quadrature position operator   $\hat{X}_M=\frac{\hat{b}+\hat{b}^\dagger}{\sqrt{2}}$ where $\hat{a}$ ($\hat{b}$) is the annihilation operator for the optical (mechanical) mode. 
The mode index $L,M$ will be set as $1,2$ in the simulation. 
The unitary evolution operator by this Hamiltonian is given as $\hat{U}_{k,l}[T_{k,l}]=\exp[\ii T_{k,l}\hat{n}_L^{k}\hat{X}_M^{l}]$ where $T_{k,l}$ is the dimensionless strength.
The challenge lies in both nonlinear regimes at the level of a few energy quanta and the deep-strong coupling that largely dominates over free oscillations of both optical and mechanical modes.   
We focus on \old{the lowest orders to demonstrate feasibility}  with $k=1$ and $l=1$ representing deep-strong coupling in \old{movable cavity mirror} systems causing radiation-pressure-like coupling linear in the mechanical operator and the $l=2$ representing the membrane-in-the-middle optomechanics (Fig. \ref{fig:scheme}a).

\subsection{Dispersive-Rabi gate method}

\begin{figure}[]
\includegraphics[width=\textwidth]{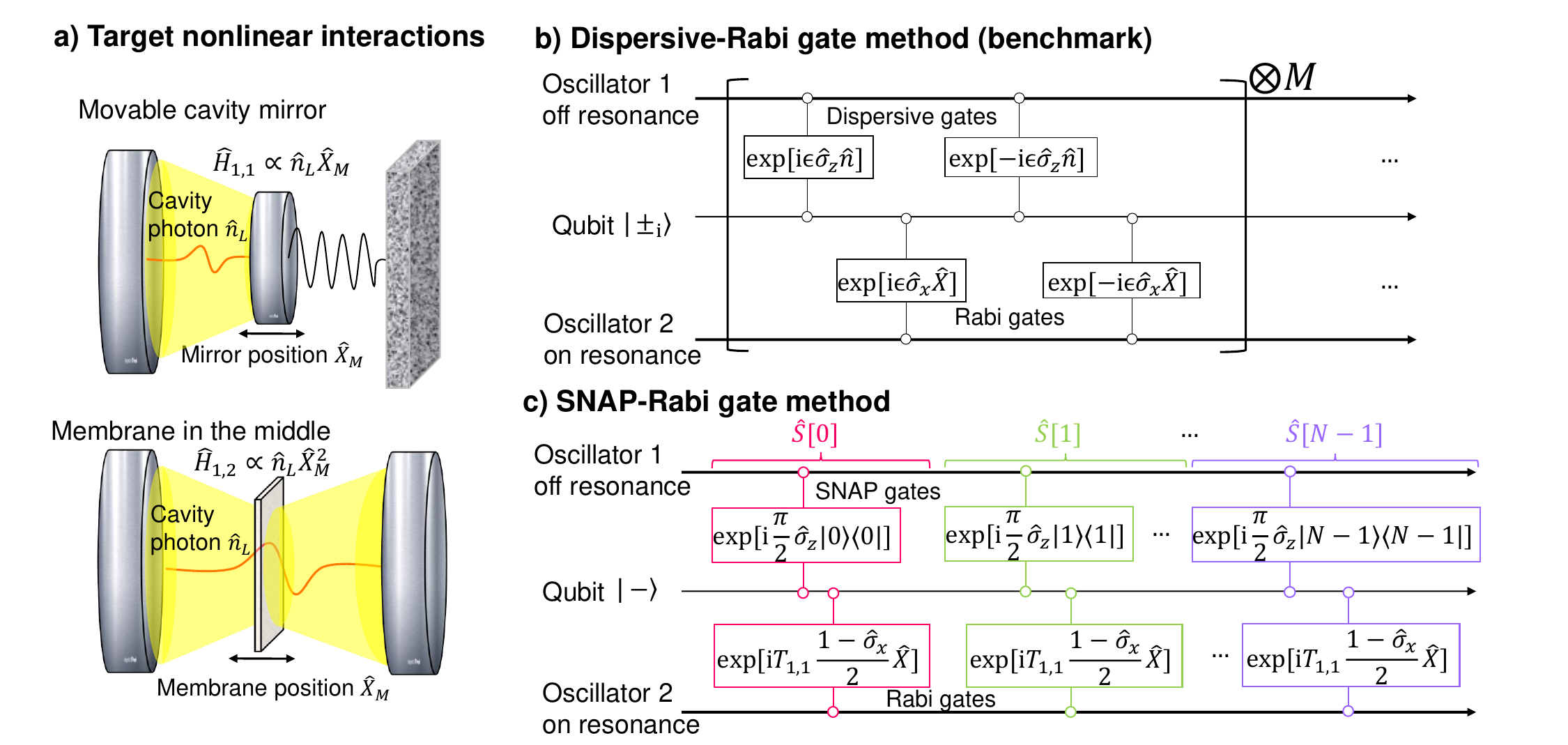}
\caption{
Quantum simulation circuits of optomechanical interactions \old{switching between two types of} qubit-oscillator gates. 
\textbf{a} Target nonlinear interactions involving  radiation pressure \old{coupling} in movable cavity mirror $\hat{H}_{1,1}$ or membrane-in-the-middle coupling $\hat{H}_{1,2}$ between an optical cavity mode and a mechanical mode. 
We aim to simulate these systems using physical experimental setups such as trapped ion or circuit QED \old{by mapping cavity and mechanical modes to two harmonic oscillators}.
\textbf{b} Circuit diagram illustrating the incremental \old{dispersive-Rabi gate} approach \new{serving as a benchmark}, alternating dispersive (off-resonant) and Rabi (on-resonant) \old{qubit-oscillator} interactions to induce an \old{effective radiation-pressure} coupling \old{between the modes}.
\old{This sequence should be repeated multiple times $M\gg 1$ to simulate a strong target coupling accurately.}
\textbf{c} Circuit diagram showing the non-incremental \old{SNAP-Rabi gate approach} using $N$ photon-number selective dispersive gates \old{(SNAP gates)} and  Rabi gates to accurately reproduce the movable cavity mirror Hamiltonians. 
\old{In both methods, the membrane-in-the-middle coupling can be simulated by substituting $\hat{X}$ by $\hat{X}^2$.}
} \label{fig:scheme}
\end{figure}


Our first approach, \new{which serves a benchmark to a new method in Sec. II. B,} considers a qubit that  interacts with two oscillators in the different regimes of resonance alternatingly \old{between dispersive and Rabi qubit-oscillator gates}, mediating a virtual effective  interactions between the oscillators deterministically without  the need for any  measurement, as in Fig. \ref{fig:scheme}b. 
\old{The basic premise is that by switching between these two different coupling regimes and leveraging the non-commuting Pauli operators $\hat{\sigma}_j$ with $j=\mathrm{x,y,z}$, the resulting operator transformations can progressively build up higher-order nonlinear interactions.}
This circuit \new{takes advantage of} the Baker–Campbell–Hausdorff (BCH) formula~\cite{Van-Brunt2015}\old{, which \new{allows the multiplication of} non-commuting operators to be approximated as a series of commutators. 
This formula possesses the power to systematically generate higher order nonlinearity when applied recursively}~\cite{ParkNJP2018,KalajdzievskiQuantum2021CVDecomp,JacobsPRL2007}.
It first builds a tripartite interaction, where a qubit non-demolitionally  mediates the indirect effective interactions between \old{the} two oscillators. 
A previous version of this \old{incremental} approach exploiting a mediating qubit has been employed for various purposes ~\cite{ParkSciRep2017Kerr,Cubic2017,TufarelliPRA2009Qubit}. 
\new{The BCH  formula can approximate the method in \cite{JacobsPRL2007} for small parameters, reducing it to the dispersive-Rabi gate method.}


For example, alternating two different kinds of unitary interactions such as dispersive gates $\hat{U}_z=\exp[\ii \epsilon\hat{\sigma}_z \hat{n}_1]$ \cite{Eickbusch2021SupConDispersive}   and Rabi gates $\hat{U}_x=\exp[\ii \epsilon\hat{\sigma}_x \hat{X}_2]$ \cite{NoriRabiRMP2019, Forn-DiazRMP2019Rabi} \old{as in the following sequence makes a weak bosonic interaction:}
 \begin{align}
 \hat{O}_{1,1}[2\epsilon^2]=\hat{U}_x \hat{U}_z \hat{U}_x^\dagger \hat{U}_z^\dagger\approx \exp[-2\ii\epsilon^2 \hat{\sigma}_y  \hat{n}_1\hat{X}_2] 
 \label{eq:incre}
 \end{align}
for a very small $\epsilon\ll 1$.
The qubit degree of freedom can be ignored if we initially prepare the eigenstate of the engineered Pauli operator $\hat{\sigma}_y$, such as $\ket{\pm_\ii}$.
The dispersive interaction can arise from the Rabi coupling far from resonance~\cite{Zueco2009}, and can be linearized to Rabi intereaction by applying strong displacement operations~\cite{vanLoock2008HybridOptics}. 
Recent advancements in the experimental realization of  these interactions have been made  in cavity~\cite{AuffevesPRL2003RydbergDispersiveHaroche, Friebe2018} and circuit QED \cite{Premaratne2017,peronnin19,TouzardPRL2019Dispersive,Eickbusch2021SupConDispersive} and trapped ion systems~\cite{FluhmannHomeNature2019,FluhmanPhysRevLett2020Trappedion}.
However, this incremental approach  has limited validity at a high strength,  requiring a large number $M\gg1$ of repetitions to achieve a high strength target gate \old{\cite{ParkNJP2018,Van-Brunt2015,KalajdzievskiQuantum2021CVDecomp}}.
\old{The total resource usage in terms of total coupling strength is counted as $4M \epsilon$.}
We can improve the approximation using concatenated operations as in Appendix \ref{sec:improving}, and online squeezing~\cite{Hastrup2020squeezing}. 
\old{A more complex circuit using concatenated BCH operations can} create $\hat{O}_{k,l}[T_{k,l}]$ that simulates  ideal operations $\hat{U}_{k,l}[T_{k,l}]$ for higher $k$ and $l$ as \old{theoretically shown} in \cite{Cubic2017}.

\subsection{SNAP-Rabi  gate method}

To overcome the limitations of the incremental \new{method}, we  introduce a non-incremental approach \new{that leverages} SNAP gates \cite{SchusterNature2007, HeeresPRL2015} to address each Fock state independently.
We utilize a circuit that bears resemblance with \cite{kudraPRXQ2022robust}.
For approximate simulation, we use \old{SNAP gates} that  flip the qubit selectively for a \new{specific} Fock state $\ket{n}$.  
Combined with Rabi gates, the following sequence constructs a selective optomechanical coupling on $n$-th Fock subspace $\hat{\Pi}_n\equiv\ket{n}\bra{n}$:
\begin{align}
    &\hat{S}[n]\equiv \exp[-\ii \frac{\pi}{2} \hat{\Pi}_n
    \hat{\sigma}_z]\exp[\ii T_{1,1} n \frac{\hat{1}+\hat{\sigma}_x}{2}\hat{X}_2]\exp[\ii \frac{\pi}{2} \hat{\Pi}_k \hat{\sigma}_z]\nonumber\\
    &=(1-\hat{\Pi}_n)(\ket{+}\bra{+}\exp[\ii T_{1,1} n \hat{X}_2]]+\ket{-}\bra{-})+\hat{\Pi}_n (\ket{+}\bra{+}+\ket{-}\bra{-}\exp[\ii T_{1,1} n \hat{X}_2]]),
    \label{eq:seloptcoup8}
\end{align}
\old{where $\ket{\pm}$ are $\hat{\sigma}_x$ eigenstates}.
Repeating \new{this sequence} for all $n$ \old{upto a finite number $N$} \new{yields} an $N$-th order approximation of the optomechanical interaction \old{without incremental limitations}:
\begin{align}
\hat{\mathbb{S}}_{1,1}[T_{1,1}]&\equiv\prod_{n=1}^N \hat{S}[n].
\label{eq:repeatedselec}
\end{align}
In the qubit $\ket{-}$ subspace, \cmmnt{we see that the finite-order approximation of order $N$ is achieved with} the erroneous operation \old{is} eliminated by orthogonality. 

An improved SNAP-Rabi gate sequence \new{in terms of simulation accuracy and resource efficiency is given as} (Fig. \ref{fig:scheme}c):
\begin{align}
   &\mathbb{S}_{1,1}[T_{1,1}]= \exp[-\ii \frac{\pi}{2}\sum_{n=0}^{N-1}\hat{\Pi}_{n} \sigma_z]\prod_{n=0}^{N-1} \overbrace{ \exp[\ii T_{1,1} \frac{\hat{1}-\hat{\sigma}_x}{2}\hat{X}_2]\exp[\ii \frac{\pi}{2}\hat{\Pi}_{n} \sigma_z]}^{\hat{S}[n]}\nonumber\\
   &=\ket{-}\bra{-}(\hat{1}-\sum_{n=0}^N \hat{\Pi}_n)\exp[\ii N T_{1,1} \hat{X}_2]+\ket{+}\bra{+}(\hat{1}-\sum_{n=0}^N \hat{\Pi}_n)+ \ket{-}\bra{-}\sum_{n=0}^N \exp[\ii n T_{1,1} \hat{X}_2]\hat{\Pi}_n+ \ket{+}\bra{+}\sum_{n=0}^N \exp[\ii (N-n) T \hat{X}_2]\hat{\Pi}_n
\end{align}
\old{where the term $\exp[-\ii \frac{\pi}{2}\sum_{n=0}^{N-1}\hat{\Pi}_{n} \sigma_z]$ can be omitted for resource efficiency by sacrificing the accuracy.
A visual depiction of how the SNAP-Rabi gate sequence constructs the simulated optomechanical interaction in a progressive, selective manner is provided in Appendix \ref{sec:disgram}.
} 
This gate uses only total strength $T_{1,1}N/2$  of the Rabi gates, \new{as opposed} to the $T_{1,1}N(N+1)/4$ in the previous approach in (\ref{eq:repeatedselec}), and thus is even more resource-efficient for a large $N$.
 \old{The total resource usage of total coupling strength is given as $T_{1,1} N/2+\pi N$.}
It also yields better fidelities under qubit errors, due to the absence of an large erroneous displacement depending on $N(N+1)$ in (\ref{eq:repeatedselec}) and a better behavior in the high Fock subspace beyond $\hat{\Pi}_N$.

\new{It is noteworthy that} by using a Rabi interaction of the form $\exp[\ii t  F(n)\frac{\hat{1}+\hat{\sigma}_x}{2}\hat{X}_2]$ with a modulated strength \new{function $F(n)$}, \old{for example $F(n)=n^{k>1}$}, instead of a standard Rabi interaction in (\ref{eq:seloptcoup8}), we can simulate an arbitrary-order optomechanical interaction. 
If we modulate the Rabi gate strength to implement a second order optomechanical interaction, we \old{obtain:}
\begin{align}
   &\mathbb{S}_{2,1}[T_{2,1}]= \exp[-\ii \frac{\pi}{2}\sum_{n=0}^{N-1}\hat{\Pi}_{n} \sigma_z]\prod_{n=0}^{N-1} \exp[\ii (2n+1) T_{2,1} \frac{\hat{1}-\hat{\sigma}_x}{2}\hat{X}_2]\exp[\ii \frac{\pi}{2}\hat{\Pi}_{n} \sigma_z].
\end{align}
The total Rabi gate usage is given as $T_{2,1} N^2/2$, in contrast to $O[N^3]$ order total strength in the previous method.
In general, implementation of $\mathbb{S}_{k,1}[T_{k,l}]$ uses Rabi gates with total strength $O[N^k]$, in contrast to the previous method that uses $O[N^{k+1}]$.
\old{The relative resource scaling advantages of the SNAP-Rabi approach compared to other methods is analyzed in more depth in Appendix \ref{sec:scaling}.}

To engineer a higher optomechanical interaction of the form \old{$\hat{\mathbb{S}}_{1,2}[T]$ in a similar manner}, we require access to a second order Rabi interaction of the form $\exp[\ii t \hat{\sigma}_x \hat{X}^2]$ with a \old{varied} strength $t$.
The standard Rabi gate can be transformed into the second order Rabi gate of the form $\hat{\sigma}_x \hat{X}^2$ at large detuning, with a noise-controlled squeezing occurring concurrently, as observed in ~\cite{SanduRJP2015DispersiveRabi}.
Alternatively, the second  \old{and higher} order Rabi interaction can be efficiently engineered using a decomposition method \cite{ParkNJP2018}.
\new{By employing higher-order Rabi interactions engineered via decomposition methods or derived from dispersive interactions at large detuning, our technique can simulate the membrane-in-the-middle interaction $\hat{U}_\mathrm{1,2}$ and more generally $\hat{U}_\mathrm{k,l}$, combining such interactions with SNAP gates as outlined in (\ref{eq:seloptcoup8})}.
This SNAP-Rabi technique can be extended further to multipartite gates such as $\exp[\ii \chi \hat{n}_1^{k_1}\hat{n}_2^{k_2}\hat{X}_3^{k_3}\hat{X}_4^{k_4}..]$, providing a versatile toolbox for multimode bosonic simulations.
The following sections compare these simulated \old{dynamics by engineered} optomechanical interactions with the benchmark Kerr-based method.
\new{The novel SNAP-Rabi gate method  provides exponential improvements in both accuracy and resource usage beyond those offered by benchmarks.
This method can be also adapted to engineer a cross-Kerr gate by simultaneously using SNAP gates on both modes.
}

\section{Results and discussion}
\label{sec:simulation}

 The \old{discussion} in this section focuses on \old{assessing} the non-Gaussian properties of the simulated \old{nonlinear dynamics} that are not accessible by the linearized \old{Gaussian one}.
Additionally, we compare our methods with an approach based on  \old{naturally existing} higher-order \old{Kerr} nonlinear interactions, which is briefly described in Appendix. \ref{append:Benchmark}.
 \old{This Kerr-based method also needs many repetitions by $M$ times, due to its incremental limitation.}
 
In our system, the implemented optomechanical \old{gates} $\hat{O}_{k,l}[T_{k,l}]$ from \old{incremental dispersive-Rabi method} (\ref{eq:incre}) \old{and $\hat{\mathbb{S}}_{k,l}[T_{k,l}]$ from \old{non-incremental SNAP-Rabi methods} (\ref{eq:repeatedselec})} are not identical to the ideal optomechanical interactions $\hat{U}_{k,l}[T_{k,l}]$, and therefore, it is crucial to investigate the extent to which \old{the observable effects} can reproduce the \old{target nonlinear dynamics} for \old{various} input states, especially at \old{deep-strong coupling regime} $T_{k,l}\approx 1$. 
Nonlinear optomechanical \old{processes} allow the phase-insensitive generation of coherent phase-sensitive effects (displacement and squeezing) on \old{the mechanical mode}, which is \old{not possible with} linearized versions. 
Therefore, phase-insensitive coherent states $\rho_\mathrm{prc}[\beta]=\int_{\phi=0}^{2\pi} d\phi \ket{\beta \ee^{\ii\phi}}_1\bra{\beta \ee^{\ii\phi}}=\ee^{-|\beta|^2}\frac{\beta^{2 \hat{n}}}{\hat{n}!}$ with Poissonian statistics and thermal states $\rho_\mathrm{th}[\bar{n}]=\left(\frac{\bar{n}}{\bar{n}+1}\right)^{\hat{n}}/(\bar{n}+1)$ with Bose-Einstein statistics are chosen as input states, whose mean photon number $\bar{n}$  can be experimentally adjusted.
\new{A thermal state represents the maximum entropy average of arbitrary pure states, subject to the constraint of a given average number of quanta.
}
These \old{classical} states can be easily \old{prepared} \old{for any oscillators,} providing experimentally straightforward initial states  to \old{test} quantum effects \old{from the simulations}.
For entanglement analysis of \old{simulated interaction}, we use phase-sensitive coherent states to understand the impact of amplitude and phase noise. 
The state in the on-resonant mode simulating  mechanical mode is chosen to be in a \old{ground} state or in thermal states with the same average quanta from realistic experimental considerations.

\begin{figure}[]
\includegraphics[width=0.9\textwidth]{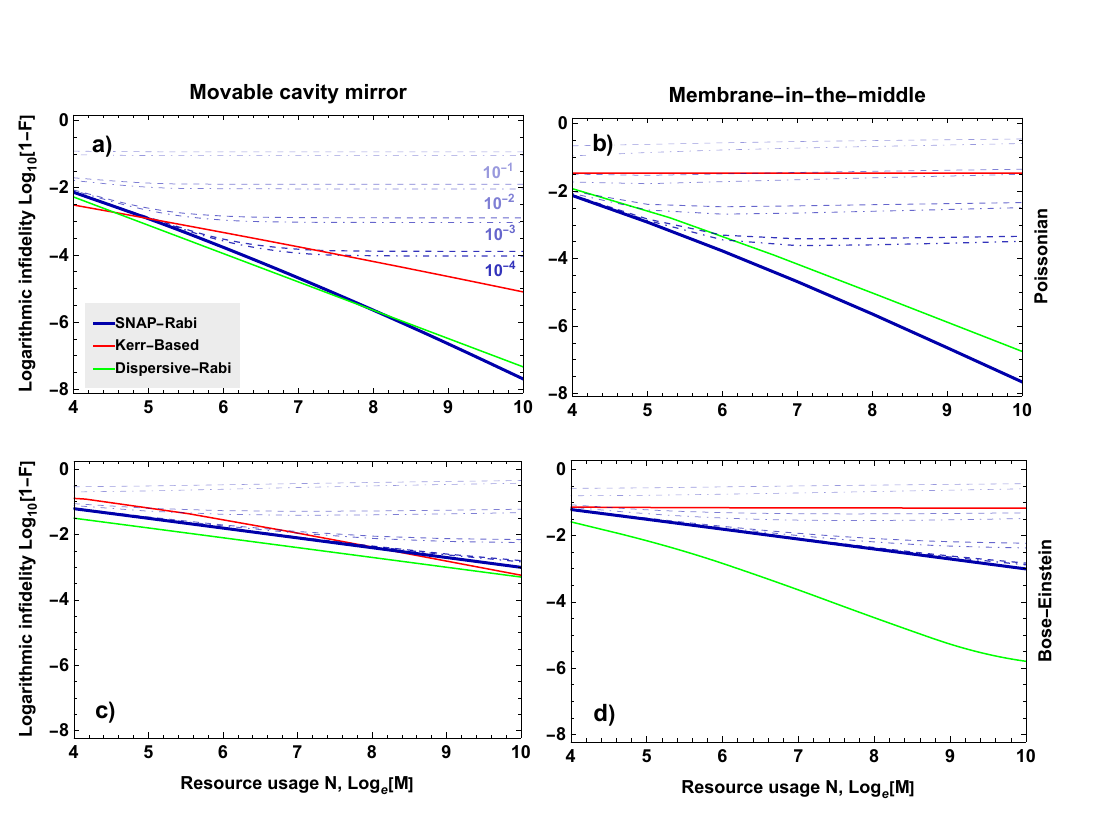}
\caption{ Simulation infidelity  versus resource usage of various methods. 
 Logarithmic infidelity vs. \old{simulation order} $N$ \old{proportional to the resource used} in the \old{SNAP-Rabi} method \old{and logarithmic repetition number $\log M$ used in the other methods } for \old{\textbf{a,b} Poissonian noise and  \textbf{c,d} Bose-Einstein noise both with $\bar{n}=1$} at \old{\textbf{a,c} movable cavity mirror dynamics at strength} $T_{1,1}=1$ \old{and \textbf{b,d} simulation of membrane-in-the-middle dynamics at $T_{1,2}=1$}. 
Different curves \old{show the impact}   of qubit loss (dashed)  and dephasing errors (dot-dashed)  during the Rabi gates.
} \label{fig:fidelity}
\end{figure}

We now assess the \old{accuracy of the simulated optomechanical interactions} based on three important criteria: 
\old{i) fidelity between output states of ideal and simulated interactions,
ii) ability to reproduce optomechanical phenomena like noise-driven displacement and squeezing,
iii) accuracy in generating target entanglement.}
The \old{rough evaluation} is \old{given by} the fidelity, defined as $F=\mathrm{Tr}[\sqrt{\sqrt{\rho_\mathrm{id}}\rho_\mathrm{re}\sqrt{\rho_\mathrm{id}}}]^2$,  between the output states by the \old{target} ideal interactions $\rho_\mathrm{id}$ and the approximate interactions $\rho_\mathrm{re}$.
This quantifies how close the output states from the simulated dynamics are to the ideal target dynamics and serves  as the \old{average} measure of closeness of the states. 
Higher fidelity indicates the simulation more accurately reproduces the desired nonlinear effects.
In Fig.~\ref{fig:fidelity}, the SNAP-Rabi gate method on Poissonian and Bose-Einstein noise shows a linear scaling in the resource usage for the infidelity drop, corresponding to an exponential improvement over the incremental dispersive-Rabi and Kerr-based methods that show exponential scaling in the resource usage for the infidelity drop.
For example, at $N=10$ of the SNAP-Rabi gate, the total resource gate strength consumed is about $0.5\%$ of that by the dispersive-Rabi with similar fidelity. 
This exponential enhancement in resource usage is maintained even for simulating the higher-order membrane-in-the-middle interaction. 
The results demonstrate the significantly higher accuracy and resource-efficiency of the SNAP-Rabi gate approach in reproducing the target nonlinear dynamics.
The fidelity scaling trends and analysis of how the different simulation methods react to increased resources are further detailed in Appendix \ref{sec:scaling}.
\new{In our stability analysis, we considered the effects of losses and noise. 
Each Rabi and SNAP gate was subjected to a constant level of qubit dephasing ($dq$), boson loss ($d\eta$), and qubit loss ($dr$), although in practical scenarios these values could depend on gate strengths. 
These losses led to a saturation in accuracy for large $N$, a consequence of the interplay between enhanced approximations and accumulating errors.
Notably, these losses had a more substantial impact on the benchmark methods (dispersive-Rabi method and Kerr-based method) due to their significantly higher resource usage compared to the SNAP-Rabi gate method. A detailed summary of the effect of boson loss is provided in Appendix \ref{sec:bosonloss}.
}

Secondly, we investigate optical noise-induced mechanical quantum coherence, which is a basic nonlinear effect at a low number of quanta.  
The nonlinear optomechanical interaction can induce displacement \old{in the movable cavity mirror}.
It is  quantified by    the signal-to-noise ratio (SNR) of displacement \old{comparing the mean induced displacement to the initial uncertainty quantifying how well the simulated interaction displaces the mechanical state beyond Gaussian noise} (see Appendix \ref{sec:snr} for its definition). 
The signal-to-noise ratio (SNR) compares the induced displacement of the mechanical state to its initial uncertainty. 
A higher SNR indicates the nonlinear optomechanical interaction can displace the mechanical state beyond the original noise levels.
Squeezing defined by  the smallest eigenvalue of a covariance matrix can be induced in the mechanical mode by membrane-in-the-middle interaction from phase-insensitive states in the optical mode depending on the \textit{phase-insensitive} photon number. 
The mechanical quantum \old{Gaussian} coherence arising from the noise in the optical mode  serves as a key \old{effective witness}  of the genuine nonlinear optomechanical  interaction at the quantum level.
The \old{noise-induced} squeezing \old{of the membrane-in-the middle} also appears  qualitatively different  from many other methods used to induce squeezing ~\cite{SetePRA2014SqueezingCEM}. 
\old{The accuracy of the SNR and induced squeezing has a similar resource scaling as the fidelity.}
\old{The details about these metrics is explained in Appendix \ref{sec:snr}.}

In Fig~\ref{fig:x2p} a), we demonstrate that the SNRs for the \old{Poissonian} and  \old{Bose-Einstein noise} with different $\bar{n}$ values in the \old{simulated} optical mode, obtained \old{from the simulated movable cavity mirror interactions} $\hat{O}_{1,1}[T_{1,1}]$ and $\hat{\mathbb{S}}_{1,1}[T_{1,1}]$, approach those from the ideal  optomechanical interaction $\hat{U}_{1,1}[T_{1,1}]$ for various $T_{1,1}$, especially for a low $\bar{n}$. 
In Fig.~\ref{fig:x2p} b), we illustrate the minimum variance of quadratures \old{by various processes simulating $\hat{U}_{1,2}[T_{1,2}]$}. 
These results demonstrate  the squeezing generated beyond a Gaussian approximation \old{of linearized interaction}, as indicated by the minimum variance falling below shot-noise limit  $\Delta X_p^2<1/2$, which implies the existence of a nonclassical phase-sensitive state. 
For both the \old{Poissonian} and \old{Bose-Einstein noises} in the \old{simulated} optical mode  with a \old{ground} state in the \old{simulated} mechanical  mode  at the initial time,  the general trends of  the SNRs and induced squeezing by the SNAP-Rabi gate  are \old{highly overlapping} with that by the ideal processes even at a \old{deep-strong} strength beyond incremental regime, suggesting that our approach effectively achieves the desired nonlinear effect  \old{with a higher accuracy than the Kerr-based approach}. 

\begin{figure}[thp]
\includegraphics[width=1.0\textwidth]{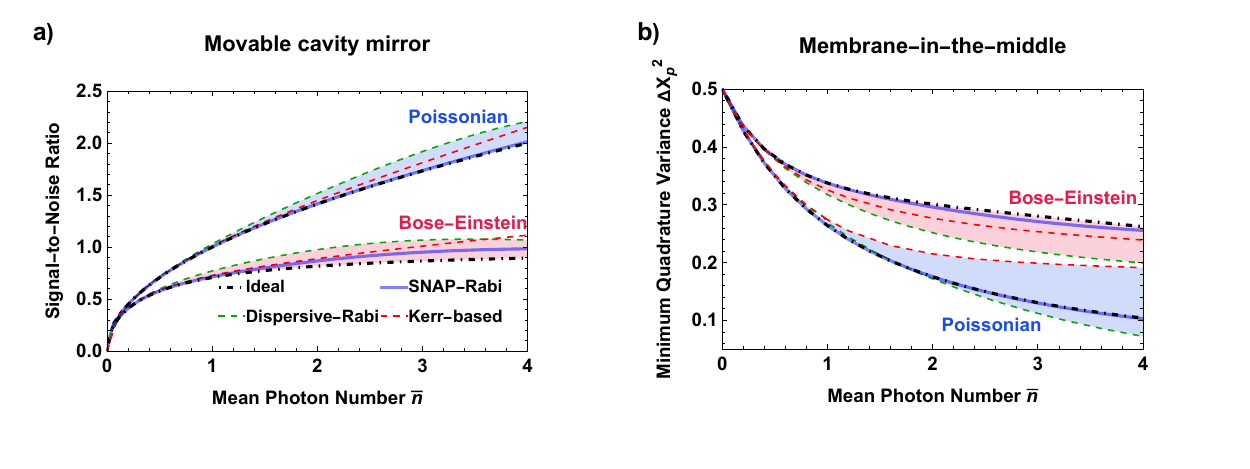}
\caption{  
Signal-to-noise ratio (SNR) of displacement and squeezing in the simulated mechanical mode induced by the ideal and \old{simulated dynamics}. 
\textbf{a} SNR of the displacement in the movable cavity mirror. 
The incremental dispersive-Rabi gate was made of $M=50$ repetitions of the unit setups, Kerr-based gate was made of $M=10$ repetitions, and SNAP-Rabi gate approach utilized $N=10$ approximation order.
The target strength was set as $T_{1,1}=1$.
\textbf{b} Minimum quadrature variance $\Delta X_p^2$  induced by the ideal \old{and approximate} membrane-in-the-middle dynamics at $T_{1,2}=1$. 
\old{Repetition of $M=50$ was used for incremental dispersive-Rabi method, while Kerr-based approach was without repetition.}
Variance below the shot noise limit (0.5) indicates the generation of squeezing and nonclassical mechanical states.
In all cases, the \old{SNAP-Rabi gate method} shows excellent agreement with ideal results, validating its ability to accurately reproduce \old{the target dynamics}.
} \label{fig:x2p}
\end{figure}

\begin{figure}[thp]
\includegraphics[width=\textwidth]{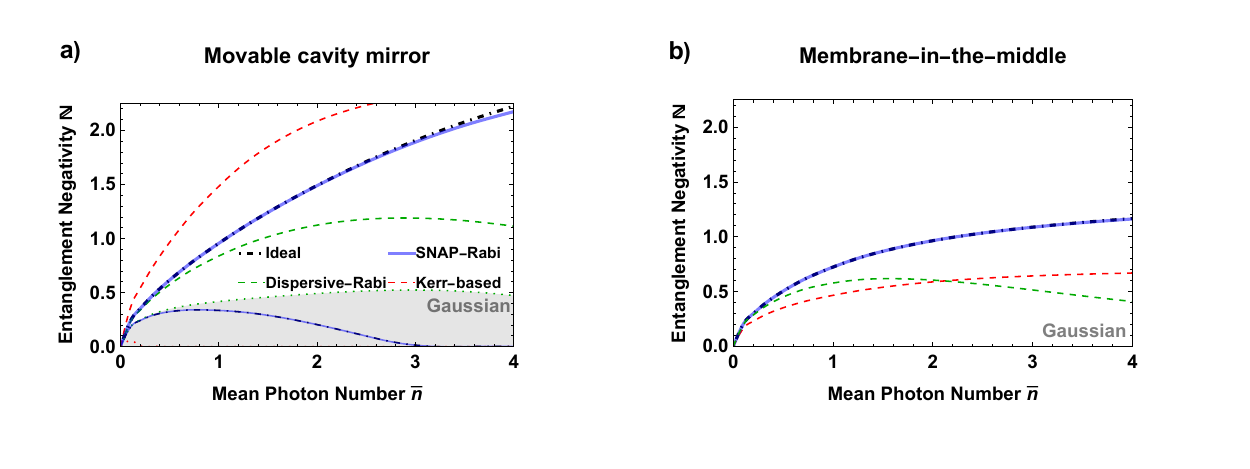}
\caption{ \old{Entanglement} negativity  and Gaussian negativities  generated by the ideal and \old{simulated} dynamics.
The simulation parameters are the same as in Fig. \ref{fig:x2p}.
\old{Entanglement above 1 ebit ($N_c=1/2$) and Gaussian entanglement (gray shade) indicates presence of non-qubit and non-Gaussian entanglement.}
In \textbf{b}, the output state has purely non-Gaussian entanglements due to the inherent non-Gaussian nature of the interaction, as indicated by  $0$ Gaussian negativities. 
In all cases the output state has a Schmidt rank $\infty$ demonstrating the continuous variable nature of the entanglement.
The \old{SNAP-Rabi gate} method closely reproduces both the Gaussian and non-Gaussian entanglement of the ideal interaction. 
This demonstrates its ability to simulate the complex hybrid continuous-discrete variable entanglement dynamics.
  } \label{fig:neg}
\end{figure}

The \old{third} criterion characterizes \old{the non-local aspect of} various dynamics via entanglement negativity. 
We quantify the entanglement of a bipartite state density operator $\rho$ \old{of simulated optical and mechanical modes} by the entanglement negativity  $\mathbb{N}[\rho]=\frac{\mathrm{Tr}[|\rho^\mathrm{PT}|]-1}{2}$, where $\rho^\mathrm{PT}$ is the partial transpose of $\rho$  and $\mathrm{Tr}[|\cdot|]$ denotes the trace norm \cite{neg,neg2}. 
Entanglement negativity quantifies the amount of entanglement between the optical and mechanical modes.
Higher negativity indicates greater quantum entanglement.
Note that bipartite qubit entanglement of the form $\ket{\psi_\mathrm{qub}}\equiv2^{-1/2}(\ket{1}_1\ket{0}_2\pm \ket{0}_1\ket{1}_2)$ has a maximal negativity of $\mathbb{N}[\ket{\psi_\mathrm{qub}}\bra{\psi_\mathrm{qub}}]=1/2$ or 1 ebit.
\old{This exceeds the entanglement predicted by the covariance matrix under Gaussian approximation}.  
On the other hand, continuous variable states such as Gaussian states  can have more than 1 ebit of entanglement, but they lack quantum non-Gaussian properties.
Entanglement exceeding 1 ebit  with  non-Gaussian characteristics \old{beyond the covariance matrix} belongs to  a completely distinct class. 
An ideal optomechanical interaction has a \old{specific entangling} nature due to the \old{product} of two  different operators,  $\hat{n}_L$ and $\hat{X}_M$, which have discrete and continuous spectra of eigenvalues. 
Generating such hybrid entanglement is a key signature of nonlinear quantum effects. 

In Fig.~\ref{fig:neg}, the entanglement negativity  is compared to Gaussian entanglement negativity,  \old{which refers to the entanglement predicted by the covariance matrix under Gaussian approximation,} produced by the ideal and simulated dynamics. 
\old{Both the simulated interactions and ideal optomechanical coupling generate entanglement that exceeds the Gaussian entanglement negativity, indicating the presence of authentic non-Gaussian entanglement beyond what is predicted within the Gaussian approximation framework.}
States diagonal in the Fock basis, such as a \old{Bose-Einstein} and a \old{Poissonian noise}   in \old{simulated} optical mode, cannot generate entanglement and only result in classical correlations.
\old{In the simulated optical mode, t}herefore, we investigate input coherent states with various average photon number $\bar{n}$ that lacks quantum \old{nonclassicality and} non-Gaussianity. 
Both \old{target} ideal and \old{simulated} \old{dynamics} generate entanglement exceeding $1$ ebit. 
This entanglement surpasses the entanglement of any Gaussian states with the same covariance matrix~\cite{Laurat2005Gaussian}, demonstrating \old{both} authentic non-Gaussian, \old{beyond-two-qubit} entanglement.
The output state by the membrane-in-the-middle dynamics has  non-Gaussian entanglements without any \old{entanglement predicted by the covariance matrix within Gaussian approximation}.
The agreement of the \old{SNAP-Rabi gate with the target ideal coupling} is much closer \old{than} both the \old{dispersive-Rabi gate} and the Kerr-based \old{simulation}.
\old{The exponential enhancement in resource-scaling of the accuracy in the simulation of the entanglement negativity is maintained.}
This result demonstrates that SNAP-Rabi gate method reproduces  the complex continuous-discrete variable entanglement arising from the nonlinear coupling \old{more precisely} than the Kerr-based method \old{and the dispersive-Rabi gate method}.
\old{The intensity-intensity correlation between the optical and mechanical modes is analyzed for the different simulation methods and summarized in Appendix \ref{sec:non-local}, providing further validation of the approaches.}

\new{
Our work primarily operates within the interaction picture and omits the effective frequency. 
In current experimental optomechanics setups, the effective mechanical frequency ($\omega_m$) typically falls within the MHz to GHz range. 
The actual experimental optomechanical interaction strength ($g_m$), however, is typically much smaller, with $g_m/\omega_m$ on the order of $10^{-3}$ to $10^{-6}$.
In our simulation, we can achieve an ultrastrong coupling where the resource coupling (denoted as $T/2$) is on the order of $1$, such that $T_{1,1}/\omega_m$ reaching a value of approximately $2$. 
This results in a simulated optomechanical interaction strength that is enhanced by several orders of magnitude relative to current state-of-the-art experiments.
This significant enhancement of the interaction strength underscores the potential of our proposed method to advance research in this field. 
The SNAP-Rabi gate approach demonstrates a substantial efficiency advantage over conventional techniques such as the dispersive-Rabi and Kerr-based methods. 
This considerable gain in resource efficiency allows the SNAP-Rabi method to accurately simulate deep-strong nonlinear optomechanical interactions that remain difficult to directly probe and investigate in existing experimental optomechanical systems.
It allows for the exploration of deep-strong nonlinear couplings between bosonic oscillators, a domain that remains challenging to access with current technology.
}

\section{Conclusions}

\old{Quantum simulation of nonlinear bosonic interactions in challenging deep-strong coupling can allow proof-of-principle tests of many new phenomena. 
 }
We \old{introduced} an \old{accurate} and efficient method for  \old{simulating} \old{challenging} nonlinear dynamics \old{by alternating qubit-oscillator coupling regimes}. 
\old{Our technique does not rely on existing bosonic nonlinearities and provides an exponential improvement in accuracy resource efficiency over previous methods.}
Our simulations demonstrate that the \old{simulated dynamics} exhibit the essential nature of a \old{new} nonlinear \old{generator of coherence and nonclassical states and hybrid} entangler, validating the accuracy of our scheme in reproducing the target dynamics in a previously unexplored regime.
The signal-to-noise ratio, and entanglement negativity analysis confirm the ability  to reproduce \old{the complex dynamics} of ideal optomechanical systems beyond the accuracy of existing Kerr nonlinearity-based approaches. 
The SNAP-\old{Rabi  gate} method based on SNAP gates \cite{SchusterNature2007, HeeresPRL2015}, in particular, shows excellent agreement with ideal optomechanical interactions.
The proposed approach can be extended to simulate higher-order nonlinearities by utilizing higher-order Rabi gates \new{or dispersive gates}, broadening the class of dynamics that can be simulated.

These \old{first proposals focused on specific targets} provide \old{broad and } valuable tools for hybrid continuous-discrete variable quantum information processing and simulations of nonlinear bosonic systems. 
However, implementing this method \old{needs further experimental development}, including switching a single qubit between resonant and dispersive coupling with multiple cavity modes, a capability \old{never tested} in current \old{superconducting} circuit QED and ion trap systems limited to fixed interaction regimes. 
While \old{advancements in achieving switchable qubit-oscillator couplings on platforms like superconducting circuits suggest that an experimental demonstration may be feasible with continued progress}~\cite{Zueco2009,GaoPRX2018,Eickbusch2021SupConDispersive,SchoelkopfNature2020,PanPRX2022}, more research is required to fully explore their capabilities \old{and overcome limitations in current systems regarding fixed interaction regimes}. 
Our analysis serves as an important step toward accessing intriguing dynamics for future quantum technologies.

\section*{Acknowledment}
K.P. and R.F. acknowledge the grant No. 22-27431S. 
P.M. acknowledges the European Union’s HORIZON Research and Innovation Actions under Grant Agreement no. 101080173 (CLUSTEC). 
We also acknowledge EU H2020-WIDESPREAD-2020-5 project NONGAUSS (951737) under the CSA - Coordination and Support Action.


\appendix
\counterwithin {figure} {section}
\section{Incremental dispersive-Rabi gate method}
\label{sec:improving}

\old{The incremental approach utilizes the commutation relations of the Pauli operators to systematically generate higher-order nonlinear interactions through operator transformations.
By alternating unitary operations containing Pauli operators with opposite signs, undesired terms can be suppressed via cancellations. However, the validity of the incremental approximations is limited for high interaction strengths, necessitating many repetitions of weak gate operations.
}
The basic construction of operators for the incremental method can be achieved by a complex arrangement \old{of two types of interactions}:
\begin{align}
&\exp[-\ii t_1\hat{A} \hat{\sigma}_y] \exp[\ii t_2\hat{B} \hat{\sigma}_z]\exp[2\ii t_1\hat{A} \hat{\sigma}_y]\exp[\ii t_2\hat{B} \hat{\sigma}_z]\exp[-\ii  t_1\hat{A} \hat{\sigma}_y]\nonumber\\
&=\exp \left[-\ii \left(2 t_1\hat{A} +\frac{\pi }{2}\right) \sigma _y \exp[2 \ii t_2\hat{B}  \hat{\sigma}_x]-\ii\frac{ \pi}{2}   \sigma _y\right]\nonumber\\
&\approx\exp[\ii \sin[2t_2\hat{B}]\cos[2 t_1\hat{A}]\hat{\sigma}_z+\ii\sin[4t_1A]\sin[B t_2]^2\hat{\sigma}_y]\nonumber\\
&\approx\exp[\ii 2t_2\hat{B}\cos[2 t_1\hat{A}]\hat{\sigma}_z], \label{eq:concat}
\end{align}
and using a sequence with different signs as
\begin{align}
&\exp[-\ii  t_1\hat{A} \hat{\sigma}_y] \exp[\ii  t_2\hat{B} \hat{\sigma}_z]\exp[2\ii t_1\hat{A} \hat{\sigma}_y]\exp[-\ii  t_2\hat{B} \hat{\sigma}_z]\exp[-\ii  t_1\hat{A} \hat{\sigma}_y]\nonumber\\
&=\exp \left[2 \ii \hat{A} t_1 \sigma _y \left(\cos[2 \hat{B} t_2] \exp[-2 \ii
   \hat{A} t_1 \sigma _y]-\ii \sigma _x \sin[\hat{B} t_2]\right)\right]\nonumber\\
&=\exp \left[2 \ii \hat{A} t_1 \sigma _y \exp[-2 \ii
   \hat{A} t_1 \sigma _y]\left(\cos[2 \hat{B} t_2] -\ii \exp[2 \ii
   \hat{A} t_1 \sigma _y] \sigma _x \sin[\hat{B} t_2]\right)\right]\nonumber\\
&=\exp \left[2 \ii \hat{A} t_1 \sigma _y \exp[-2 \ii
   \hat{A} t_1 \sigma _y]\exp[-\ii\hat{B} t_2\exp[2 \ii
   \hat{A} t_1 \sigma _y] \sigma _x]\right]\nonumber\\
&\approx\exp[\ii \sin[2t_2\hat{B}]\sin[2 t_1\hat{A}]\hat{\sigma}_x-\ii \sin[4 t_1 A ] \sin[t_2B ]^2\hat{\sigma}_y]\nonumber\\
&\approx\exp[\ii 2 t_2\hat{B}\sin[2 t_1\hat{A}]\hat{\sigma}_x]. \label{eq:odd}
\end{align}
where the last line was obtained in a weak strength limit $t_1,t_2\ll 1$.
Here,  a product form of $\hat{B}f(\hat{A})$ with an infinite order polynomial $f$ has been achieved in the exponent, which can become the sources of high-order nonlinear interactions.

This approximation can be further improved by suppressing the errors by concatenating with similar operations with opposite signs by which the unwanted terms are cancelled out.  
More rigorously, we can use the Fourier series expansion to obtain the target interaction efficiently. Using $\exp[-\ii t_1\hat{A} \hat{\sigma}_y]\hat{\sigma}_z \exp[\ii t_1\hat{A} \hat{\sigma}_y]=\cos[2  t_1\hat{A}]\hat{\sigma}_z+\sin[2  t_1\hat{A}]\hat{\sigma}_x=\hat{\sigma}_z\exp[\ii\hat{\sigma}_y 2t_1 \hat{A}]$ \old{and $\exp[-\ii t_1\hat{A} \hat{\sigma}_y]\hat{B} \exp[\ii t_1\hat{A} \hat{\sigma}_y]=\hat{B}$}, we have:
\begin{align}
&\exp[-\ii t_1\hat{A} \hat{\sigma}_y]\exp[\ii t_2\hat{B}\hat{\sigma}_z ]\exp[\ii t_1\hat{A} \hat{\sigma}_y]=\exp[\ii t_2\hat{B}\hat{\sigma}_z\exp[\ii 2t_1 \hat{A}\hat{\sigma}_y]]. \label{eq:braiding}
\end{align}
Note that this is an exact equation, the validity of which is not limited by the parameters of $t_1, t_2$. Now these operations can be combined to make the target function $\exp[\ii T\hat{\sigma}_\theta\hat{A}\hat{B}]$ for certain strength $T$ and a certain Pauli matrix $\hat{\sigma}_\theta$. 
Let us assume for simplicity that $2t_1$ takes discrete values, such that $2t_1=k \tau$ with  $k=...,-2,-1,0,1,2,...$ and $t_2$ can take arbitrary values. Then we are given the operations as $\exp[\ii t_2\hat{B}\hat{\sigma}_z\exp[\ii k \tau \hat{A}\hat{\sigma}_y]]$.   Here, we use the Fourier series expansion $x=\ii\sum_{k=1,2,...}\frac{(-1)^k}{k}\exp[\ii k x]-\frac{(-1)^k}{k}\exp[-\ii k x]$, to have 
\begin{align}
&\prod_{k=1,2,...}^{k_\mathrm{max}=\infty}\exp[\ii t_2\hat{B}\hat{\sigma}_z\frac{(-1)^k}{k}\exp[\ii k \tau \hat{A}\hat{\sigma}_y]]\exp[-\ii t_2\hat{B}\hat{\sigma}_z\frac{(-1)^k}{k}\exp[-\ii k \tau \hat{A}\hat{\sigma}_y]]\nonumber\\
&\approx \exp[\ii t_2\hat{B}\hat{\sigma}_z\left(\sum_{k=1,2,...}^{k_\mathrm{max}}\frac{(-1)^k}{k}\exp[\ii k \tau \hat{A}\hat{\sigma}_y]-\frac{(-1)^k}{k}\exp[-\ii k \tau \hat{A}\hat{\sigma}_y]\right)]=\exp[-\ii t_2\hat{\sigma}_x  \tau \hat{A}\hat{B}].
\end{align}
For a finite order approximation with $k_\mathrm{max}$, we can get a fairly good approximation as the strength required is \old{sufficiently} weak for a high order contributions. 


A better approximation can be obtained using the following approximation: $A B\approx 0.6 (-\ee^{\ii (-A-B)}+\ee^{\ii (A-B)}+\ee^{-\ii (A-B)}-\ee^{\ii (A+B)})$.
Therefore, we obtain
\begin{align}
&\exp[-\ii t_2\hat{\sigma}_y  \tau \hat{A}\hat{B}]\approx \exp[-\ii t_2\hat{\sigma}_y (-\ee^{\ii \hat{\sigma}_x (-A-B)}+\ee^{\ii \hat{\sigma}_x (A-B)}+\ee^{-\ii \hat{\sigma}_x (A-B)}-\ee^{\ii \hat{\sigma}_x(A+B)})]\nonumber\\
&=\exp[\ii t_2\hat{\sigma}_y \ee^{\ii\hat{\sigma}_x (-A-B)}]\exp[-\ii t_2\hat{\sigma}_y \ee^{\ii \hat{\sigma}_x (A-B)}]\exp[-i t_2\hat{\sigma}_y \ee^{-i \hat{\sigma}_x(A-B)}]\exp[i t_2\hat{\sigma}_y \ee^{i \hat{\sigma}_x(A+B)}].
\end{align}
Here we used $\hat{\sigma}_x^2=1$. Here, each term can be engineered from (\ref{eq:braiding}) with substitution $A\rightarrow A\pm B, B\rightarrow 1$. 

We can go a step further by using Fourier transform corresponding to a continuous limit of Fourier series expansions. We note that for any quantity $x$, $\int \mathrm{d}x  x \exp[\ii x p]=-\ii \delta'(p)$ by the Fourier transform with a integration variable $p$ where $\delta'(p)$   is the derivative of Dirac delta function.
Now noticing the approximation $\delta(p)=\lim_{a\rightarrow 0}\frac{\exp[-(p/a)^2]}{\sqrt{\pi}a}$, we can approximate $\delta'(p)\approx -\frac{2 p \ee^{-\frac{p^2}{a^2}}}{\sqrt{\pi } a^3}$. This implies the following approximation $x\approx \ii\int \mathrm{d}p   \frac{2 p  \ee^{-\frac{p^2}{a^2}}}{\sqrt{\pi } a^3} \exp[-\ii x p]$. Now arbitrarily choosing small $a$ and substituting integration by summation by setting $\mathrm{d}p=a$, we have $x\approx \ii  \sum_{k}   \frac{2 k  \ee^{-k^2}}{\sqrt{\pi } a} \exp[-\ii k a x ]$.
Using this equation, we will get 
\begin{align}
&\exp[-\ii t_2\hat{\sigma}_x  \tau \hat{A}\hat{B}]=\exp[-\ii t_2\hat{\sigma}_z   \hat{B} i\int \mathrm{d}p   \frac{2 p  \ee^{-\frac{p^2}{a^2}}}{\sqrt{\pi } a^3} \exp[-\ii \tau\hat{\sigma}_y p \hat{A} ]]\approx\exp[\ii t_2\hat{B}\hat{\sigma}_z\left( \ii  \sum_{k}   \frac{2 k  \ee^{-k^2}}{\sqrt{\pi } a} \exp[-\ii \hat{\sigma}_y k a \tau \hat{A} ]\right)]\nonumber\\
&\approx\prod_{k} \exp[\ii t_2\hat{B}\hat{\sigma}_z \ii    \frac{2 k  \ee^{-k^2}}{\sqrt{\pi } a} \exp[-\ii \hat{\sigma}_y k a \tau \hat{A} ]].
\end{align}

Alternatively, we can enhance the strength of the achieved gates by applying the additional squeezings~\cite{AlbarelliPRA2018}, which transforms a quadrature operator as $S[-r]\hat{X}S[r]=\hat{X} \ee^r$.  
This transformation  can also be applied to  operators containing quadrature operators $\exp[f(\hat{X})]$ with an arbitrary function $f(\cdot)$. 
In our analysis, we focus on the simplest case of $S_2[-r]\exp[\ii t \hat{n}_1 \hat{X}_2]S_2[r]=\exp[\ii t \ee^r \hat{n}_1 \hat{X}_2]$, which shows that squeezing enhances the gate strength from $t$ to $t \ee^r$. 
On the other side, we can also enhance the strength of the number operator as $S[-r]\hat{n}S[r]\approx\hat{n} \cosh[r]^2$, thus obtaining a similar enhancement by squeezing $S_1[-r]\exp[\ii t \hat{n}_1 \hat{X}_2]S_1[r]\approx\exp[\ii t \cosh[r]^2 \hat{n}_1 \hat{X}_2]$. 
This technique can increase the feasibility of experimental schemes and overcome the limit of weak strengths of individual operations, although it cannot enhance the non-Gaussian nature of the interaction.


\section{Benchmarks based on higher order interactions}
\label{append:Benchmark}

An alternative approach to achieve a high strength \old{radiation-pressure} interaction $\hat{U}_{1,1}$ is to linearize a naturally arising higher-order cross-Kerr interaction by a large displacement, as demonstrated in \cite{YanagimotoPRL2020Cubic, ZhouOptExp2021Simulation}. 
Specifically, we can use the following sequence of interactions:
\begin{align}
   \exp[-\ii \frac{\A^2}{2}t \hat{n}_1] \exp[\ii \A \hat{P}_2]\exp[\ii t \hat{n}_1\hat{n}_2]\exp[-\ii \A \hat{P}_2]\exp[-\ii t \hat{n}_1\hat{n}_2]\approx\exp[\ii t\A \hat{n}_1\hat{X}_2].
\end{align}
This approximation linearizes one interacting mode, and holds well when $t\ll 1$, and $\alpha\gg 1$.
\old{Again, repetition of this sequence by $M$ times can increase the strength of the simulated gate or the accuracy of the simulation.}
Here, the Kerr interaction can also be engineered by setting $A=\hat{n}_1$ and $B=\hat{n}_2$ from two dispersive interactions as in (\ref{eq:incre}), or we can use the naturally existing Kerr interaction existing in QED systems~\cite{LiaoPRA2020OptoMechKerrSim, ZhouOptExp2021Simulation}.
Another approach to achieve $\hat{U}_{1,1}$ is to transform a Fredkin gate by a strong drive as described in \cite{Yin2021}, the process of which is simply described as :
\begin{align}
\exp[\ii t \hat{a}^\dagger \hat{a}(\hat{b}^\dagger \hat{c}+\hat{c}^\dagger \hat{b})]\ket{\xi}_c\approx \exp[\ii t \hat{a}^\dagger \hat{a}(\hat{b}^\dagger \xi+\xi^* \hat{b})]\ket{\xi}_c.
\end{align}
However, we do not delve into  this approach, as it requires an even less accessible three-mode interaction than the cross-Kerr interaction.
We can also simulate the \old{membrane-in-the-middle} interaction $\hat{U}_{1,2}$ using Kerr interactions and squeezing.
This can be achieved using the following transformation:
\begin{align}
    \exp[- \ii \frac{t}{2} (\ee^{2r}-1)\hat{n}_1]\exp[-\ii t \ee^{2r}\hat{n}_1\hat{n}_2]S_2[r]\exp[\ii t \hat{n}_1\hat{n}_2]S_2[-r]\approx \exp[\ii t \sinh[2r]\hat{n}_1\hat{X}_2^2]
\end{align}
\old{where the squeezing operator on the mechanical mode $S_2[r]=\exp[-\frac{r}{2} \hat{b}^{\dagger2}+\frac{r}{2} \hat{b}^{2}]$.}
This approximation holds well when $t\ll 1$ \old{and $|r|\gtrsim 1$}.
These Kerr-interaction-based methods are inefficient in terms of resource usage due to the necessity of inverse interactions, but are the main targets for comparison with our protocols.
\new{A similar technique was used to enhance the strengths of Kerr gate in \cite{Bartkowiak2014}.}

\section{SNAP-Rabi gate sequence}
\label{sec:disgram}

\begin{figure}[]
\includegraphics[width=0.58\textwidth]{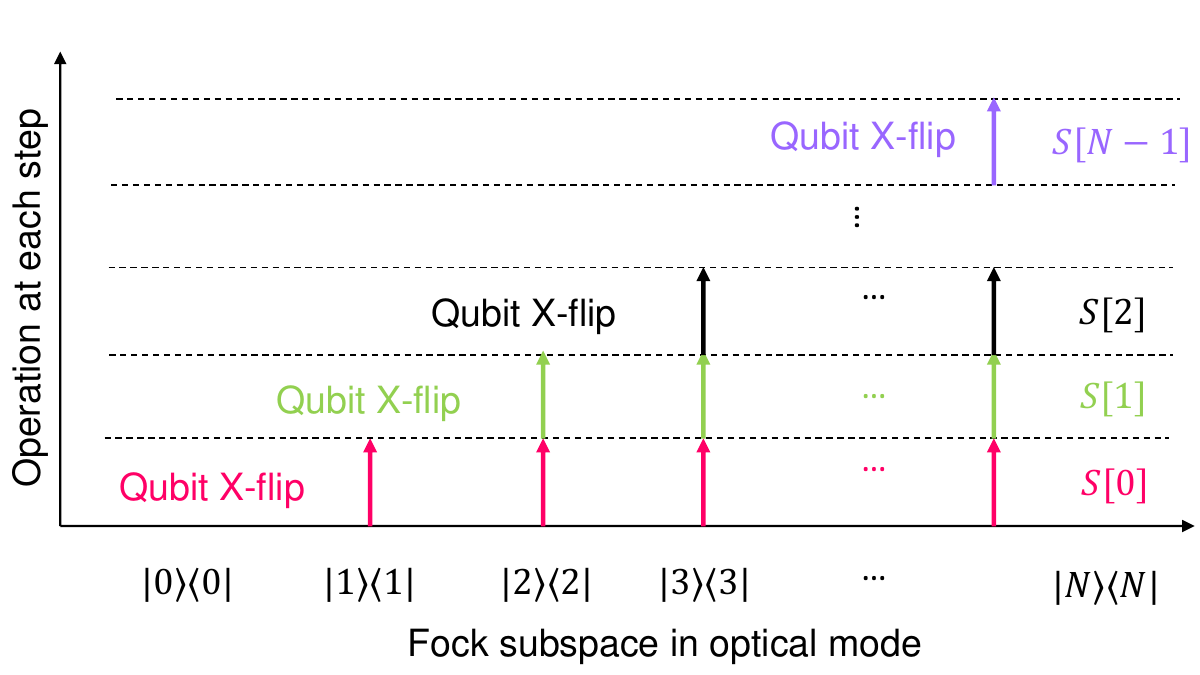}
\caption{\old{A diagram visualizing the SNAP-Rabi gate sequence on joint qubit-oscillator states.
Each horizontal level shows the effect of the gate set $S[n]$.
It flips the qubit in X-basis for $n$-th Fock space, and displaces the mechanical mode conditioned by the higher Fock spaces.
The upward arrows represent conditional displacement $D[\ii T/\sqrt{2}]$ of the mechanical oscillator.
This induces selective displacements that build up the simulated nonlinear optomechanical interaction in each Fock subspace.}
} \label{fig:diagram}
\end{figure}

\old{
Fig. \ref{fig:diagram} describes intuitively how SNAP-Rabi gate builds up the simulated movable cavity mirror dynamics.
By combining conditional qubit flips mediated by the SNAP gates and conditioned displacements mediated by the Rabi gates, the sequence builds up the simulated nonlinear interaction in a selective, progressive manner within each Fock subspace.
Overall, the total displacement applied to the mechanical mode is proportional to the Fock number $n$ in the optical mode, as required to accurately simulate the optomechanical dynamics.
}

\section{Signal to noise ratio and induced mechanical squeezing}
\label{sec:snr}

\old{In addition to fidelity, other metrics can also characterize how well the simulated interactions reproduce the target optomechanical dynamics.}
The  first-order optomechanical interaction $\hat{U}_{1,1}$  possesses phase sensitivity in the mechanical mode, which results in displacement beyond the noise that is proportional to the photon number in the optical mode.  
In contrast, the linearized interaction cannot generate net mechanical displacement for phase-insensitive states, such as PRC or thermal states.  
 Therefore,  the successful implementation of the optomechanical interaction $\hat{U}_{1,1}[T]$ can be demonstrated by producing a mechanical displacement beyond the linearized regime, which generates  displacement beyond shot noise and results in a mechanical state with non-zero off-diagonal elements in the Fock state basis in the mechanical mode.
These off-diagonal elements indicate the quantum coherence induced by the optical noise.

To evaluate the effectiveness of implementing the optomechanical interaction $\hat{U}_{1,1}[T]$, we can use the SNR to measures the  displacement generated relative to the initial Gaussian uncertainty \old{located at the phase space origin}. 
In the Heisenberg picture, any operator $\hat{O}$ under a unitary evolution $\hat{U}$ \old{is described as} $\hat{U}^\dagger \hat{O}\hat{U}$.
  The momentum operator shifted by the adjoint action of an ideal $\hat{U}_{1,1}$ interaction is described by
\begin{align}
 &\hat{P}'_2
 =\hat{P}_2-t \hat{n}_1,~~\delta \hat{P}_2=\hat{P}'_2-\hat{P}_2=-t \hat{n}_1. 
 \label{eq:shiftmom}
\end{align}
In contrast, the momentum operator shifted by the linearized interaction is described by:
\begin{align}
 &\hat{P}'_2
 =\hat{P}_2-t \hat{X}_1,~~\delta \hat{P}_2=\hat{P}'_2-\hat{P}_2=-t \hat{X}_1. 
 \label{eq:shiftlinear}
 \end{align}
 The \old{average} momentum shift in (\ref{eq:shiftmom}) can be non-zero for an  input Fock state in the optical mode $1$, while that in (\ref{eq:shiftlinear}) is zero due to the phase insensitivity of the Fock states, or the states diagonal in the Fock basis used in the simulation.
 \old{Therefore, a non-zero average momentum shift can serve as  evidence of the nonlinear optomechanical interaction, but a sufficient strength interaction is required to have the shift beyond shot-noise limit.}

We can calculate the SNR using the formula $\old{\mathrm{SNR}=}\av{\delta \hat{P}_2}/\sqrt{\mathrm{Var}(\delta \hat{P}_2)}$ \old{for various average photon number $\bar{n}$ in the optical mode, where the average $\av{\cdot}$ and the variance $\mathrm{Var}(\cdot)$ is calculated for the initial states}. 
 Initially, the simulated mechanical mode is assumed to be in a vacuum state without quantum coherence in any basis, including the Fock state basis.
For PRC in the optical mode, the SNR is calculated to be equal to $\old{\mathrm{SNR}^\mathrm{(PRC)}=}\sqrt{\bar{n}}=\alpha$ for the ideal interaction $\hat{U}_{1,1}[T]$, which is the coherent amplitude of the state, and is independent of the strength $T$. 
For a thermal state in the optical mode, the SNR increases approximately as $\mathrm{SNR}^\mathrm{(th)}=\sqrt{\frac{\bar{n}}{\bar{n}+1}}\stackrel{\bar{n}\ll 1}{\to}\sqrt{\bar{n}}$, and  asymptotically saturates to the value $1$ when $\bar{n}\gg 1$. 
\old{Again, the SNR is independent of the strength $T$.}



%

On the other hand, a higher order interaction such as  $\hat{U}_\mathrm{1,2}[T]=\exp[\ii T \hat{n}_1\hat{X}_2^2]$ can induce both the quantum coherence and the nonclassical squeezing effects. 
The presence of a square term in $\hat{X}$ in the Hamiltonian provides a mechanism for quantum squeezing \old{in the mechanical mode in general}.
By using \old{this second order} ideal optomechanical interaction, we can derive  operator relations in Heisenberg picture as follows:
\begin{align}
&\hat{P}_2'
=\hat{P}_2+2T\hat{n}_1\hat{X}_2.
\end{align}
These relations allow us to obtain the momentum variance:
\begin{align}
&\Delta P_2^2=\av{P_2^2}_T-\av{P_2}_T^2\nonumber\\
&=\av{P_2^2}-\av{P_2}^2+4T \av{\hat{n}_1}\Big(\av{\frac{\hat{X}_2\hat{P}_2-\hat{P}_2\hat{X}_2}{2}}-\av{X_2}\av{P_2}\Big)\nonumber\\
&+4T^2\av{\hat{n}_1^2}(\av{\hat{X}_2}^2-\av{\hat{X}_2}^2).
\end{align}
We observe that  the uncertainty $\Delta P_2^2$ monotonically increases with $T$ due to the product of covariances of mechanical mode $2$ and photon number moments of optical mode $1$. 
This increased uncertainty in $P$ implies the existence of squeezing along some direction of quadrature in phase space. 

\section{Resource Scaling Analysis of Simulation Methods}
\label{sec:scaling}

\begin{figure}[]
\includegraphics[width=500px]{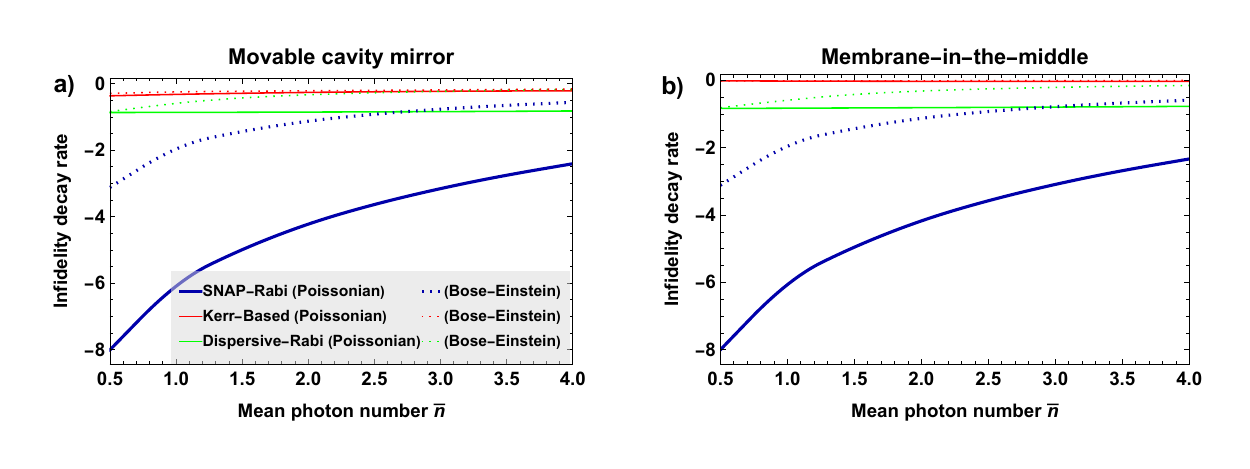}
\caption{\old{Scaling behavior against resource usage over various mean photon number of Poissonian and Bose-Einstein noise models.
\textbf{a} Simulation of movable cavity mirror dynamics.
\textbf{b} Simulation of Membrane-in-the-middle dynamics.}
} \label{fig:fidsclae}
\end{figure}

The metrics for various method of simulation can be compared in their scaling against the resources used and the mean photon number for a complete picture.
This can be quantified by the linear slope of the curves in Fig. \ref{fig:fidelity}.
The fidelities generally improve exponentially with more resource usage, modeled empirically as $1-F[R]=(1-F[0])e^{\delta_R R}$, where $R$ represents the resource used, and $\delta_R$ is termed as the infidelity decay rate representing how simulation accuracy reacts the resource usage.
In Fig. \ref{fig:fidsclae}, the infidelity decay rate for movable cavity mirror and membrane-in-the-middle dynamics are shown, with Poissonian and Bose-Einstein noise models. 
The resource usage was counted as $N$ for SNAP-Rabi gate method and $M$ for the other methods.
The SNAP-Rabi gate method shows a much larger decay rate, meaning the input resource rapidly  improves the simulation fidelity.
Noticeably the decay rate of SNAP-Rabi gate method depends more on the mean photon number of the noise model due to its finite dimension nature in Fock subspace.
Again, the Kerr-based method has a very small decay rates, meaning its repetition is not very effective in improving the simulation accuracy.

\new{Superposition states of a few Fock states can exhibit nonlinear optomechanical effects that differ from statistical mixtures, highlighting the role of quantum coherence.
Such superpositions may be useful for generating highly non-Gaussian quantum states, although preparing and characterizing specific Fock state superpositions presents additional experimental challenges.}
Here, we test the performance of the simulated first-order optomechanical operation on randomly sampled states on the off-resonant mode $1$, and a vacuum state on the resonant mode $2$.
The complex coefficients of $c_n$ random sampled  states $\sum_n \ee^{-n/2} c_n \ket{n}$ can be sampled with a constraint $-1\le \mathrm{Re}[c_n],\mathrm{Im}[c_n]\le 1$.

The SNAP-Rabi gate method yields \old{advantageously} much higher fidelities, with an infidelity as low as $1-F=10^{-8}$ using $10$ SNAP gates  for these states with average excitation number $1$ at $T=1$.
We also observe high fidelities around $1-F\approx 10^{-4.5}$ for randomly sampled states. 
Fock states in the off-resonant modes give the optimal fidelity for the simulation performed here.

\section{Other non-local characteristics}
\label{sec:non-local}
\textit{Intensity-intensity correlation}
Intensity-intensity correlation quantifies the correlation between the photon number fluctuations in the optical and mechanical modes.
This metric directly captures the nonlocal energy transfer caused by the nonlinear coupling.
It is defined as 
\begin{align}
g^{(2)}(0)=\frac{\av{\hat{n}_1\hat{n}_2}}{\av{\hat{n}_1}\av{\hat{n}_2}},
\end{align}
where $\hat{n}_{1,2}$ are the photon number operators for the optical and mechanical modes, respectively.
For ideal optomechanical interaction, we can use the following equations
\begin{align}
    \exp[-i t \hat{n}_1\hat{X}_2]\hat{n}_1\exp[i t \hat{n}_1\hat{X}_2]=\hat{n}_1, \\
    \exp[-i t \hat{n}_1\hat{X}_2]\hat{n}_2\exp[i t \hat{n}_1\hat{X}_2]=\hat{n}_2+t \hat{n}_1 \hat{P}_2+\frac{t^2}{2}\hat{n}_1^2.
\end{align}
For an ideal optomechanical interaction on the coherent states or equivalently on the phase randomized coherent states and a vacuum state in the mechanical mode, the radiation pressure coupling results in:
\begin{align}
 \av{\hat{n}_1}=\bar{n}, ~\av{\hat{n}_2}=\frac{t^2}{2}(\bar{n}^2+\bar{n}),  ~\av{\hat{n}_1\hat{n}_2}=  \frac{t^2}{2}(\bar{n}^3+3\bar{n}^2+\bar{n}),
\end{align}
  where $\bar{n}$ is the average photon number in the optical mode.
Therefore, the intensity-intensity correlation is given as
\begin{align}
    g^{(2)}_\mathrm{Poisonnian}(0)=\frac{\bar{n}^2+3\bar{n}+1}{\bar{n}^2+\bar{n}},
\end{align}
a value independent from the strength of the interactions, and diverging at $\bar{n}\to 0$ and saturating as $\lim_{\bar{n}\to \infty}=1$.
For an initial thermal state in the optical mode, the ideal interaction gives
\begin{align}
 \av{\hat{n}_1}=\bar{n}, ~\av{\hat{n}_2}=\frac{t^2}{2}(2\bar{n}^2+\bar{n}),  ~\av{\hat{n}_1\hat{n}_2}=  \frac{t^2}{2}(6\bar{n}^3+6\bar{n}^2+\bar{n}),
\end{align}
and
\begin{align}
    g^{(2)}_\mathrm{Bose-Einstein}(0)=\frac{6\bar{n}^2+6\bar{n}+1}{2\bar{n}^2+\bar{n}},
\end{align}
also diverging at $\bar{n}\to 0$ and saturating as $\lim_{\bar{n}\to \infty}=3$.

For separable states, this quantity will give $g^{(2)}(0)=1$, and deviation from this value will witness the existence of entanglement.
Deviations from these ideal values can diagnose the accuracy of simulated optomechanical interactions. 
Calculating and comparing $g^{(2)}(0)$ for the output states of both simulated and ideal interactions could provide a good metric of how well the nonlinear coupling is reproduced.

\begin{figure}[]
\includegraphics[width=500px]{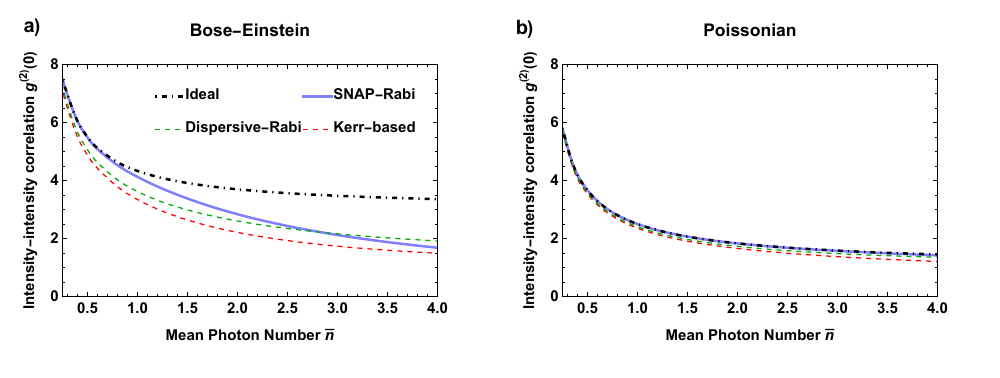}
\caption{Intensity-intensity correlation \old{$g^{(2)}(0)$} between optical mode and mechanical mode vs the average photon number $\bar{n}$ in the optical mode \old{for different simulation methods}.
\old{Initial states are } (a) an initial thermal state and (b) an initial \old{phase-randomized} coherent state in the optical mode.
The mechanical mode is assumed to be in the vacuum state.
\old{The SNAP gate method provides better agreement with the ideal optomechanical correlation, particularly for larger $\bar{n}$.}
} \label{fig:llcorr}
\end{figure}

Figure \ref{fig:llcorr} shows the intensity-intensity correlation $g^{(2)}(0)$ between the optical and mechanical modes as a function of the mean photon number in the optical mode. 
The correlation obtained from the ideal optomechanical interaction is compared against the dispersive-Rabi gate, SNAP-Rabi gate, and Kerr-based simulation methods. 
For a coherent state input, all three simulation methods can reproduce the ideal correlation relatively well across the range of $\bar{n}$, with the SNAP gate approach providing the closest match. 
This provides further evidence that the simulated dynamics mimic the desired nonlocal effects of the target nonlinear coupling.
However, for a thermal state input, simulating the ideal correlation becomes more challenging, especially in the large $\bar{n}$ limit, due to the long tail of the thermal photon number distribution.

%

\section{Boson loss}
\label{sec:bosonloss}

\begin{figure}[]
\includegraphics[width=500px]{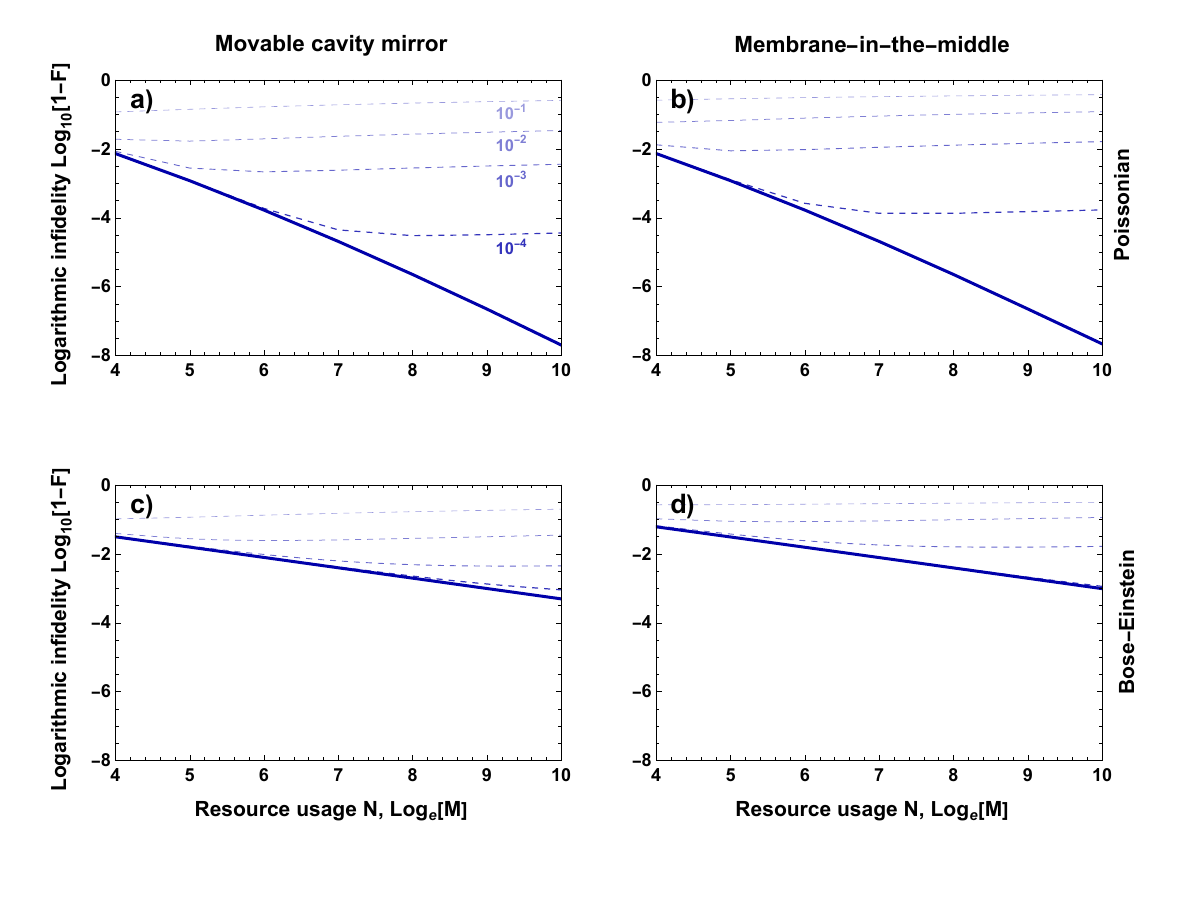}
\caption{
\new{Logarithmic infidelity versus resource usage by SNAP-Rabi gate method for simulating bosonic Gaussian operations using a movable cavity mirror and membrane-in-the-middle. 
Different lines correspond to various loss rates per gate, from $1-d\eta=10^{-4}$  to $10^{-1}$ where $d\eta$ is the loss parameter occurring at each gate. 
As resource usage increases, infidelity eventually saturates due to the constant loss rate per gate.}
} \label{fig:fidloss}
\end{figure}

\new{Bosonic loss can significantly affect the efficiency of quantum simulation protocols.
To account for these losses, we consider a loss model where each bosonic mode undergoes amplitude damping at a rate $\gamma$. 
This loss process can be described by the master equation in the Lindblad form:
\begin{equation}
\dot{\rho} = \sum_{j=1}^{2}\gamma (2\hat{a}_{j}\rho \hat{a}_{j}^{\dag} - \hat{a}_{j}^{\dag}\hat{a}_{j}\rho - \rho \hat{a}_{j}^{\dag}\hat{a}_{j}),
\end{equation}
where $\hat{a}_{j}$ ($\hat{a}_{j}^{\dag}$) is the annihilation (creation) operator for the $j$-th bosonic mode.
Any single mode density matrix $\rho$ undergoing bosonic loss channel $\Gamma_\eta$ evolves as $\Gamma_\eta[\rho]=\sum_{l=0}^\infty \frac{(1-\eta)^l}{l!} \eta^{\hat{n}/2} \hat{a}^l \rho \hat{a}^{\dagger l} \eta^{\hat{n}/2}$,  where  the loss parameter is defined as $\eta=\mathrm{e}^{-\gamma t}\in (0,1]$  and $\gamma t$ is the dimensionless damping parameter, equivalent to Kraus operator notation.

To evaluate the impact of bosonic losses on the accuracy of our simulation schemes, we calculate the infidelity of our simulation protocols under different loss rates per gate. Figure H.1 shows the logarithmic infidelity versus resource usage for the SNAP-Rabi gate method for simulating bosonic Gaussian operations using a movable cavity mirror and membrane-in-the-middle. Different lines correspond to various loss rates per gate, from $10^{-4}$ to $10^{-1}$. For both systems, the infidelity increases with loss rate and decreases with more resources used.

As for the question about dissipation in the referee report, we would like to point out that in realistic scenarios, dissipation can indeed impact the performance of our simulation protocols. 
However, our SNAP-Rabi gate method shows a certain level of resilience against bosonic losses. As can be seen from Fig. \ref{fig:fidloss}, even under moderate loss rates, the method still maintains a relatively high fidelity, indicating that it could be robust against dissipation in practical implementations. 
Moreover, the rapid decrease of infidelity with increased resource usage suggests that the detrimental effects of the mild losses can be mitigated by using more resources, even though for a larger resource usage, the infidelity saturates due to the fixed per gate loss rate.}

\bibliography{optomech}{}

\end{document}